\documentclass{desyproc}

\begin{document}
%------------------------------------
\title{Status and initial operation of ALICE}

% use the following for an author speaking on behalf of a collaboration
%
\author{{\slshape J Schukraft}  for the ALICE Collaboration\\[1ex]
CERN, 1211 Geneva, Switzerland }

% please do not modify the following 5 lines
\contribID{xy}  % will be entered by the editors
\confID{1964}
\desyproc{DESY-PROC-2010-01}
\acronym{PLHC2010}
\doi            % will be entered by the editors

\maketitle

\begin{abstract}
  After close to 20 years of preparation, the dedicated heavy ion experiment ALICE took first data with proton collisions at the LHC at the end of 2009. This article recalls the main design choices made for the detector and summarizes initial operation and performance of ALICE at the LHC; first physics results are covered elsewhere in these proceedings.
  
\end{abstract}

\section{The first 18 months: Design choices}

ALICE, which stands for A Large Ion Collider Experiment, is very different in both design and purpose from the other experiments at the LHC. Its main aim is the study of matter under extreme conditions of temperature and pressure, i.e. the Quark-Gluon Plasma, in collisions between heavy nuclei. Data taking with pp (and later p-nucleus) is required primarily to collect comparison data for the heavy ion program. However, given the specific and complementary capabilities of ALICE, a number of measurements concerning soft and semi-hard QCD processes are of interest in their own in these more elementary collisions and are part of the initial physics program~\cite{Carminati:2004fp,Alessandro:2006yt}.

Designing a dedicated heavy ion experiment in the early 90's for use at the LHC almost 20 years later posed some significant challenges: In a field still in its infancy - with the SPS lead program starting only in 1994 - it required extrapolating the conditions to be expected by a factor of 300 in energy and a factor of 7 in beam mass. The detector therefore had to be both 'general purpose' - able to measure most signals of potential interest, even if their relevance may only become apparent later - and flexible, allowing additions and modifications along the way as new avenues of investigation would open up. In both respects ALICE did quite well, as it included a number of observables in its initial menu whose importance only became clear after results appeared from RHIC (e.g. secondary vertexing for heavy quarks, particle identification up to large transverse momentum), and various major detection systems where added over time to match the evolving physics, from the muon spectrometer in 1995, the transition radiation detector (TRD) in 1999, to a large jet calorimeter (EMCAL) added as recently as 2008.

Other challenges relate to the experimental conditions expected for nucleus-nucleus collisions at the LHC. The most difficult one to meet is the extreme number of particles produced in central collisions, which could be up to three orders of magnitude larger than in typical proton-proton interactions at the same energy, and a factor two to five still above the highest multiplicities measured at RHIC. The tracking of these particles was therefore made particularly safe and robust by using mostly three-dimensional hit information with many points along each track (up to 200) in a moderate magnetic field (B = 0.5 T) to ease the problem of tracking.
In addition, a large dynamic range is required for momentum measurement, spanning more than three orders of magnitude from tens of MeV to well over 100 GeV. This is achieved with a combination of detectors with very low material thickness (to reduce scattering of low momentum particles) and a large tracking lever arm $L$ of up to 3.5 m, which gives a figure of merit for momentum resolution, $BL^2$, quite comparable to those of the other LHC experiments. In addition, the vertex detector with its six silicon planes, four with analogue read-out, can work as a standalone spectrometer with momentum and PID information to extend the low momentum range for particles that do not reach the outer tracking detectors.

And finally, Particle Identification (PID) over much of this momentum range is essential, as many phenomena depend critically on either particle mass or particle type. ALICE therefore employs essentially all known PID techniques in a single experiment, including energy loss in silicon and gas detectors, Cherenkov and transition radiation, time-of-flight, electromagnetic calorimeters, as well as topological decay reconstruction.

As the LHC luminosity with heavy ion beams is rather modest, with interaction rates of order 10 kHz or less with Pb beams, rather slow detectors can be employed like the TPC and silicon drift detectors. Only moderate radiation hard electronics and trigger selectivity are required and most of the read-out is not pipelined but uses 'track and hold'. However, because the event size in heavy ion interactions is huge (up to 100 Mbyte/event) and the statistics has to be collected in a short time (1 month/year), the DAQ has been designed for very high bandwidth of over 1 Gbyte/s to permanent storage, larger than the throughput of all other LHC experiments combined. 

The layout of the ALICE detector and its various subsystems is described in detail in~\cite{ALICEdet}.

\section{The next 18 years: R\&D and Construction}

The ALICE design evolved from the Expression of Interest (1992) via a Letter of Intent (1993) to the Technical Proposal (1996) and was officially approved in 1997.  The first ten years were spent on design and an extensive R\&D effort. Like for all other LHC experiments, it became clear from the outset that also the challenges of heavy ion physics at LHC could not be really met (nor paid for) with existing technology. Significant advances, and in some cases a technological break-through, would be required to build on the ground what physicists had dreamed up on paper for their experiments. The initially very broad and later more focused, well organised and well supported R\&D effort, which was sustained over most of the 1990's, has lead to many evolutionary and some revolutionary advances in detectors, electronics and computing~\cite{Evans:2009zz}. One example is given in the following for the 'heart' of ALICE, the Time Projection Chamber (TPC)~\cite{Alme:2010ke}.

The need for efficient and robust tracking has led to the choice of a TPC as the main tracking detector. By providing highly redundant information (up to 159 space points per track), it has to deliver reliable performance with tens of thousands of charged particles within the geometrical acceptance. In order to enhance the two-track resolution and reduce space charge distortions, a rather unusual Neon/CO$_2$ based drift gas mixture is used: the CO$_2$ reduces diffusion whereas the Neon has a low primary ionisation and large ion mobility, therefore limiting the built-up of space charge currents. The wire readout chambers are adapted to this gas with a narrow gap, as low as 2 mm, between anode wires and the pad plane. Special attention was also paid to minimise the amount of material and therefore the four cylinders of 5 m length and diameter up to 5.6 m, which make up the TPC vessel, are made of lightweight composite materials. The total amount of material traversed by a particle from the vertex through the silicon detectors to the outer part of the active TPC volume was thus kept to about 10\% of a radiation length, with the TPC operating gas a non-negligible part of the total.
The second innovation is the readout electronics:  a preamplifier/ signal shaper, operating at the fundamental thermal limit of noise, is followed by a specially developed readout chip, the ALICE TPC Read Out (ALTRO) chip. It processes digitally the signals for optimized performance at high collision rates, including a programmable digital pulse shaping circuit and zero suppression/baseline restoration algorithms.

As usual, optimisation involves compromises, and there is a price to pay for the specific choices made: The 'cool' drift gas requires a rather high drift field gradient (400 V/cm); the drift velocity depends very sensitively on temperature (which is kept constant and homogeneous to about 100 mK), pressure, electric field, and gas composition; the chambers have to be constructed with tight geometrical tolerances; the lightweight field cage easily deforms under stress or even gravitation and needs to be kept essentially force-free. In particular the drift velocity needs to be known and constantly calibrated with 
$10^{-4}$ accuracy; this is done with several independent methods including a laser system and a special drift velocity detector, while final precision is achieved after several passes using track matching with the vertex detectors.

\section{The last 18 weeks: Operation}
The very first pp collisions where observed in ALICE on November 23 2009, when the LHC slipped in, on very short notice, an hour of colliding beams for each of its four large experiments during the very early commissioning phase. Such was the penned up energy and enthusiasm about 'real data', after years of simulation exercises, that this first harvest of some 300 events, significantly less than the number of ALICE collaborators, was analysed right away and made it into a physics publication only five days later~\cite{mult09}; well before stable beams were declared on 6 December and sustained data taking could start. It took 20 years to built the experiment, one hour to take the first data, 2 days to get the first result and 3 days to finalise the author list: all of this a clear sign that physics exploitation had started for good!  

The many years of preparation, analysis tuning with simulations, and detector commissioning with cosmics during much of 2008/9 paid of quickly and handsomely with most of the detector components working with collisions 'right out of the box' and rather close to performance specifications. Within days all experiments could show first qualitative results and the first phase of LHC physics, often referred to as the 'rediscovery of the standard model', was getting under way~\cite{LHCstatrep}. The various members of the 'particle zoo' created in pp collisions made their appearance in ALICE in rapid succession, from the easy ones 
($\pi, K, p, \Lambda, \Xi, \phi,..$) in 2009 to the more elusive ones  when larger data sets were accumulated early 2010 (K*,$ \Omega$, charmed mesons, $J/\Psi$, ..). 

However, precise results and small systematic errors need more than large statistics and a good detector performance; they require a precise {\em understanding} and detector simulation as well. The next months were therefore spent on 'getting to know' the experiment in greater detail, including calibration, alignment, material distribution and detector response which are all crucial ingredients for the analysis and correction procedures.

\begin{figure}[!t]
%\centerline{\includegraphics[width=0.35\textwidth]{conversions1.eps}}
\centerline{\includegraphics[width=0.8\textwidth]{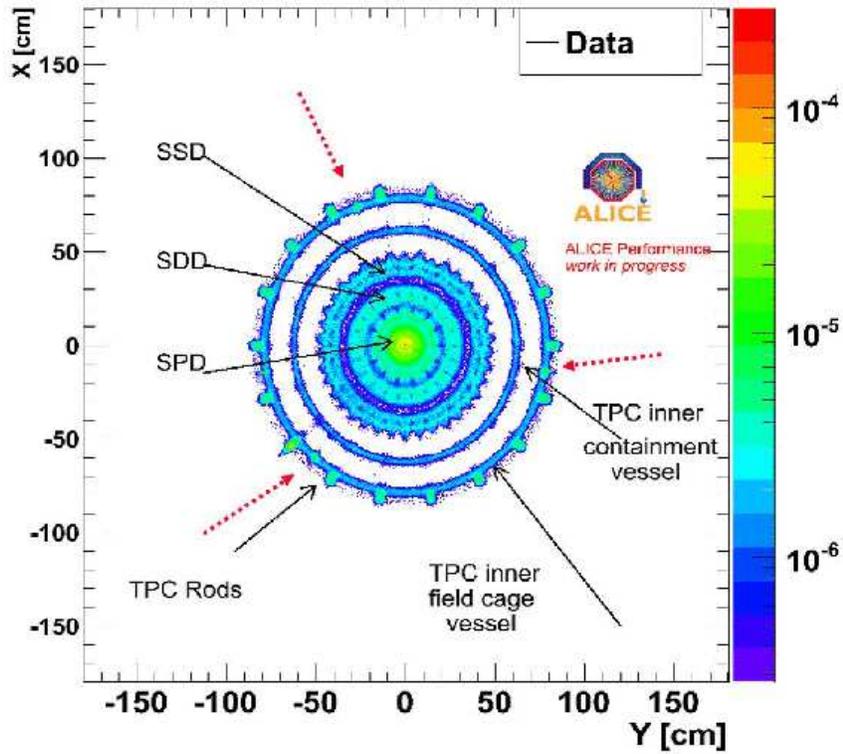}}
\caption{Reconstructed conversion points transverse to the beam direction, showing the material distribution between the vertex and the innermost two vessels of the TPC.}
\label{conversions}
\end{figure}

\begin{figure}[!t]
\centerline{\includegraphics[width=0.8\textwidth]{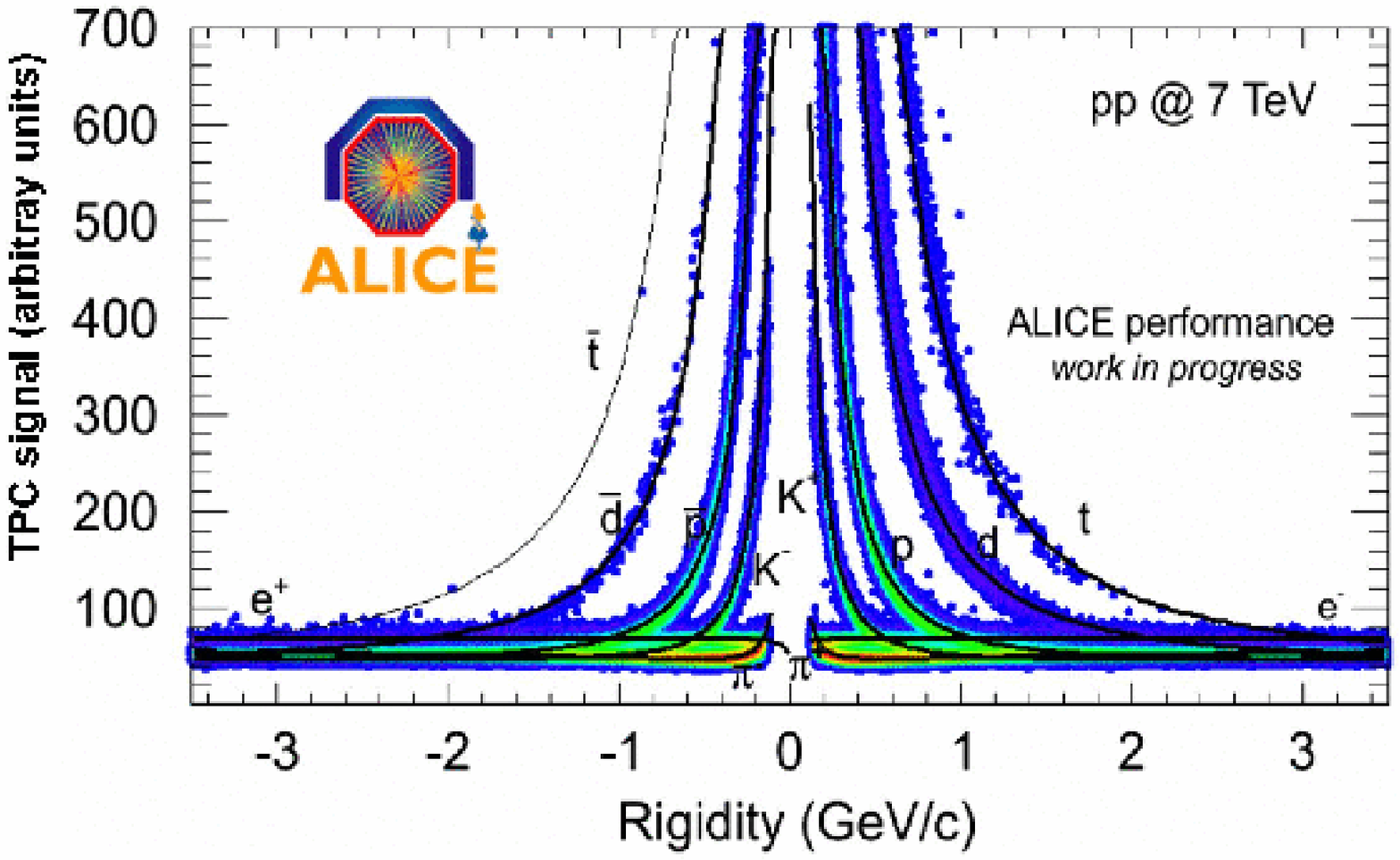}}
\caption{ Energy loss distribution versus rigidity for primary and secondary particles reaching the TPC. The lines overlaid on the distribution correspond to the expected energy loss for different particle species.}
\label{TPCdedx}
\end{figure}

An illustration of the detective work required to accurately measure the material distribution in the central part of ALICE is visible in Figure~\ref{conversions}.
It shows the distribution of reconstructed photon conversion points in a projection transverse to the beam direction; they sample in great detail the material distribution inside the detector with different structures (beam pipe, silicon detector layers, TPC vessels etc..) easily discernible. The outermost ring corresponds for example to the TPC inner field cage, with 18 'peaks' corresponding to 18 rods which support the field defining conductive aluminium strips. However, three additional peaks (marked by arrows) in the data had no known correspondence in the actual detector, and the baseline between the peaks was slightly higher than expected. After several weeks of consulting construction drawings, and, more important, the people who actually built the detector, it turned out that the additional structures were in the position where the three pieces of the field cage were joined together with a 'generous helping' of glue, whereas the increased thickness corresponded to a last minute change in thickness of the carbon fibre layers, which did not make it into the final drawings. Designing a thin detector is one thing, knowing precisely what was actually built quite another! After similar investigations in other parts of the detector, the material thickness had slightly increased overall but is now known to better than 5\% relative accuracy (i.e. 0.5\% $X/X_0$ absolute). Such accuracy is important for example for the measurement of the antiproton to proton ratio, where annihilation of antiprotons in the detector material is one of the limiting factors in reducing the systematic error~\cite{Aamodt:2010dx}.

\section{Detector status and data taking}
Most of the 18 different ALICE detector systems are fully installed, commissioned and operational, with the exception of the two systems (TRD and EMCAL) which were added more recently and are only now nearing the end of construction. Both systems have currently about 40\% of their active area installed and will be completed during the long shutdown in 2012. 

Detector alignment, which started with cosmics and continues with beam, is essentially completed for the silicon pixel (SPD) and silicon drift (SDD) vertex detectors (residual misalignment $< 10 \mu$m), and has reached about 100 $\mu$m for the SDD (where geometrical alignment and drift velocity calibration are coupled). The TPC geometry is aligned to 200-300 $\mu$m, approaching the specifications, and the outer detectors are at the mm level required for track matching. This work is still ongoing for the muon spectrometer, which could not be prealigned with cosmics given its vertical orientation along the beam line. 

Accurate gain and pulse height calibrations, which are needed in particular for the detectors used for dE/dx particle identification, is essentially complete with the TPC dE/dx resolution having reached its design value about $6\%$ for long tracks. 
The energy loss distribution in the TPC is shown in Figure~\ref{TPCdedx} versus rigidity, separately for positive and negative charges, demonstrating the clear separation between particle species reached in the non-relativistic momentum region. Note that in this plot tracks are not required to point precisely back towards the vertex and therefore many secondaries produced in the detector material are included.

The TPC drift velocity is measured precisely and continuously (with a time granularity of less than 30 min) to $< 10^{-4}$ using the collision data. The momentum resolution has reached 1\% (7\%) at 1 (10) GeV. Further calibration, in particular to correct for higher order effects of the electric and magnetic fields (ExB, local E field distortions,..), are ongoing in order to extend the accessible momentum range towards 100 GeV, where the design resolution is $< 10\%$,  including the information from the vertex detectors. Also the performance of the TOF is reaching design with a detector resolution of about 90 ps. The tight construction schedule for the electromagnetic calorimeters PHOS and EMCAL did not allow for calibration with test beams and is therefore currently done with beam data with the help of reconstructed $\pi^0$ decays.

Data taking in ALICE during 2010 will focuse on collecting a large sample ($> 10^9$) of minimum bias collisions which are needed as comparison sample for the heavy ion run later this year. By end of May, some 200 million MB events and 0.6 M events triggered with a single low $p_t$ trigger in the muon spectrometer have been recorded. The data taking efficiency is slightly above 80\%, limited somewhat by the careful and slow procedure to switch on the sensitive gas detectors after beams are brought into collisions; a procedure which was put in place as a precautionary measure during this initial LHC running. Data are automatically reconstructed shortly after data taking, and offline reconstruction as well as analysis work very satisfactory, making extensive use of the LHC computing GRID. 

After two decades of design, R\&D, construction, installation, commissioning and simulations, the ALICE experiment has 'hit the ground running' since LHC started its operation at the end of 2009. The detector is on good shape (and of the correct weight!), most systems are fast approaching design performance, and physics analysis has started and produced the first results (see elsewhere in these proceedings). While heavy ion physics will be its main subject, the collaboration has started to explore the 'terra incognita' at LHC with pp collisions, along the way gaining experience and sharpening its tools in anticipation of the first heavy ion run later this year.

% ****************************************************************************
% BIBLIOGRAPHY AREA
% ****************************************************************************

% please do not change the following line
\begin{footnotesize}

% please do not change the following line
\end{footnotesize}

% ****************************************************************************
% END OF BIBLIOGRAPHY AREA
% ****************************************************************************

\end{document}